\documentclass[pdflatex]{sn-jnl}
\usepackage{graphicx}
\usepackage{amsmath,amssymb,amsfonts}
\usepackage[title]{appendix}
\usepackage{appendix}
\usepackage{multirow}
\usepackage{natbib}
\usepackage{makecell}
\usepackage{tablefootnote}

\renewcommand{\tablename}{\textbf{Table}}
\renewcommand{\thetable}{\textbf{\arabic{table}}}
\newcommand{\rthis}[1]{\textcolor{black}{#1}}
\usepackage{float}
\begin{document}

\title[Hubble constant with profile likelihood]{Determination of Hubble constant from Megamaser Cosmology Project using Profile Likelihood}% Force line breaks with \\

\author[1,2]{\fnm{Shubham} \sur{Barua}}
 \email{ph24resch01006@iith.ac.in}%Lines break automatically or can be forced with \\
\author[2,3]{\fnm{Vyaas} \sur{Ramakrishnan}}
 \email{vyaas3305@gmail.com}
\author*{\fnm{Shantanu} \sur{Desai}}
 \email{shntn05@gmail.com}
\affil{
 Department of Physics, IIT Hyderabad, Kandi, Telangana-502284,  India}


\abstract{
The Megamaser Cosmology Project inferred a value for the Hubble constant given by $H_0=73.9 \pm 3.0 $ km/sec/Mpc. This value was obtained using Bayesian inference by marginalizing over six nuisance parameters, corresponding to the  velocities of the  megamaser galaxy systems. We obtain an independent estimate of the Hubble constant with the same data using frequentist inference. For this purpose, we use profile likelihood to dispense with the aforementioned nuisance parameters. The frequentist estimate of the Hubble constant is given by $H_0=73.5^{+3.0}_{-2.9}$ km/sec/Mpc and agrees with the Bayesian estimate to within $0.2\sigma$, and both approaches  also produce consistent confidence/credible intervals.
Therefore, this analysis provides a proof-of-principle application of profile likelihood in dealing with nuisance parameters in cosmology, which is complementary to Bayesian analysis.
}

%\noindent \textbf{Keywords:} cosmology, hubble tension, dark energy, general relativity

\maketitle

%\tableofcontents
\section{Introduction}
\label{sec:intro}
One of the most vexing problems in the current concordance $\Lambda$CDM model is the so-called Hubble tension, which refers to a $5\sigma$ discrepancy between the measurement of the Hubble constant ($H_0$) based on measurements based on the cosmic microwave background and those based on low redshift probes such as Cepheid calibrated Type Ia supernovae~\citep{tensionreview,Verde,Bethapudi,Shah21,Freedman21,DiValentino22,Verde24}. This discrepancy has been considered very seriously by the cosmology community and could be a harbinger of new Physics~\citep{KamionkowskiRiess}, although see ~\citet{Vagnozzi} for a contrarian viewpoint.

One important data analysis aspect in all recent estimates of $H_0$, is that the final value of $H_0$ has been obtained using Bayesian inference after marginalizing over other cosmological or astrophysical nuisance parameters. The marginalization is usually done using Markov Chain Monte Carlo methods~\citep{Sanjib}. In fact, almost all cosmological analyses in the last two decades have been done using Bayesian inference~\citep{Trotta08} with very few exceptions (See  ~\citealt{Hamann12,Planck13,Sarkar16} for some exceptions involving the use of frequentist inference in  Cosmology). In particle physics, on the other hand, parameter inference has always been done using Frequentist techniques, where the nuisance parameters get dispensed with using the concept of Profile Likelihood~\citep{Cowan13}.

Recently, it has been demonstrated within the Cosmology community that the process of marginalization, inherent in Bayesian analysis, could be affected by prior volume effects and can bias the final results~\citep{Adria}.
Therefore, after the advent of the Hubble tension conundrum, there has been a resurgence in the application of profile likelihood techniques to parameter inference in Cosmology~\citep{Herold22,HeroldFerreira,Campeti,Colgain24,Karwal24,Sah24} (and references therein).
In particular, \citet{Herold22} has shown that some of the null results or upper limits  on early dark energy obtained with Bayesian inference using MCMC were affected by volume effects. However,  when  profile likelihood was used to deal with the nuisance parameters, one finds $2\sigma$ evidence for early dark energy. Therefore, it behooves us to test wherever possible if the Bayesian estimates of $H_0$ agree with the frequentist estimates for some of the  $H_0$ measurements.

In this manuscript, we obtain an independent estimate of $H_0$ using data from the Megamaser Cosmology Project~\citep{Pesce} (P20, hereafter), using frequentist inference. This work is a follow-up to our recent papers on applications of profile likelihood to search for Lorentz-invariance violation using gamma-ray burst spectral lags~\citep{DesaiGanguly,Vyaas}.
For the megamaser analysis, we first independently reproduce some of the results in P20 with Bayesian inference using MCMC and then compare these results with Frequentist inference. 

This manuscript is structured as follows. We provide a brief prelude to profile likelihood and Bayesian inference in Sect.~\ref{sec:PL}. In Sect.~\ref{sec:recap}, we recap the procedure used in P20 to estimate $H_0$. We then independently reproduce these results with Bayesian inference in Sect.~\ref{sec:bayesian} and Frequentist inference in Sect.~\ref{sec:frequentist}. We then extend our analysis by allowing the matter density to be a free parameter and present the corresponding results in Sect.~\ref{sec:comparison}.  We finally conclude in Sect.~\ref{sec:conc}.

\section{Comparison of Profile Likelihood and Bayesian Inference}
\label{sec:PL}
We now compare and contrast the advantages and disadvantages of Profile Likelihood and Bayesian Inference. More details can be found in ~\citet{Herold24} (and references therein)\footnote{The profile likelihood based analysis carried out in this work is referred to as ``graphical'' profile likelihood in ~\citet{Herold24}.} along with applications to multiple cosmological problems. We shall also demonstrate its usage for the megamaser data in the forthcoming sections.

The profile likelihood is a special case of frequentist maximum likelihood analysis, where the unknown parameter vector contains only one (or very few) parameter(s) of interest along with a bunch of nuisance parameters. In the profile likelihood method, the parameter of interest is kept fixed and likelihood maximized with respect to the other nuisance parameters, and the procedure is then repeated by varying the parameter of interest. The confidence intervals are then obtained from the variation of profile likelihood as a function of parameter of interest based on the Neyman prescription~\citep{Neyman37}. In Bayesian inference, the credible intervals over the parameter of interest are obtained by marginalizing  (or integrating) the joint posterior over the nuisance parameters.  

Unlike Bayesian inference, the profile likelihood does not involve any priors. If the data are not constraining, the likelihood will be flat and the posterior will show a strong dependence on the prior. Furthermore, in Bayesian analysis since the marginalization is done in multidimensional posterior parameter space, there is a built in dependence on the prior volume. Therefore, in case the priors are not well motivated or if the posterior is not Gaussian, one can obtain unphysical results~\citep{Adria}. The impact of prior can also be more pronounced for parameters close to a physical boundary.  Recently, a large number of cosmology analyses have  found Bayesian parameter constraints to be prior dependent~\citep{Smith21,Bacon23,Pedro,Slosar23}. The likelihood used for frequentist parameter estimation is also invariant with respect to the parametrization and hence is insensitive to the aforementioned problems in Bayesian inference.

However, profile likelihood analysis also has some shortcomings. Since this  method is insensitive to the parameter volume, it prefers parameters with very small parameter volumes  and is subject to fine tuning. Also calculating frequentist confidence intervals using the full Neymann construction requires evaluation of likelihood for a large number of synthetic datasets in order to ensure correct coverage is achieved. This might not always be feasible, and hence one resorts to application of the profile likelihood  directly on the data. However, the profile likelihood method  gives the correct coverage only for a Gaussian parameter distribution or  in the asymptotic limit of a large dataset. For a given dataset and parametric model, it is possible that neither  of these two conditions are satisfied, and therefore one might not get the desired coverage.

Therefore, in light of these advantages and disadvantages of both these methods, it is important to compare the results from both these analyses.

\section{Analysis of megamaser data in P20}
\label{sec:recap} 
It has been known for a while that water megmasers in accretion disks around supermassive black holes provide very good geometric probes of distances~\citep{Hern}. The Megamaser Cosmology Project has been designed for this purpose, to find such megamaser systems and use them to measure $H_0$~\citep{Braatz08,Reid09,Reid19,Pesce20}. Such a measurement of $H_0$ is independent of distance ladder and is therefore complementary to Cepheid based and CMB-based techniques.

In 2020, P20 reported  a measurement of $H_0$ using six megamaser systems with their estimated value given by $H_0=73.9 \pm 3.0 $ km/sec/Mpc.  We now provide a brief summary of  the analysis done in the aforementioned work. 

P20 obtained precise  angular diameter distance ($D_A$) and velocity estimates to six megamaser systems. The expression for $D_A$ for a galaxy at a redshift ($z_i$) for a  flat $\Lambda$CDM cosmology is given by:
\begin{equation}
D_A  =  \frac{c}{H_0 \left( 1 + z_i \right)} \int_0^{z_i} \frac{\text{d}z}{\sqrt{\Omega_m \left( 1 + z \right)^3 + \left( 1 - \Omega_m \right)}}
\end{equation}
where $\Omega_m$ is the cosmological matter density, for which  P20 used the value of  0.315. The expected recession velocity ($v_i$) is then related to redshift by 
\begin{equation}
v_i=c z_i
\label{eq:cz}
\end{equation}
To estimate $H_0$, P20 constructed  a   likelihood given by $-0.5 \chi^2$, where $\chi^2$ is given by the following expression:
\begin{equation}
\chi^2 =  \sum_i \left[ \frac{\left( v_i - \hat{v}_i \right)^2}{\sigma_{v,i}^2 + \sigma_{\text{pec}}^2} + \frac{( D_A - \hat{D}_A )^2}{\sigma_{D,a}^2} \right] , 
\label{eqn:ChiSquared}
\end{equation}
where $\hat{v}_i$ is the measured galaxy velocity and 
$\sigma_{v,i}$ is the corresponding uncertainty; $\sigma_{\text{pec}}$ is the uncertainty in the peculiar velocity assumed to be 250 km/sec;  $\hat{D}_A$ is the distance measured from disk modeling while $\sigma_{D,a}$ denotes its uncertainty. In Eq.~\ref{eqn:ChiSquared}, the true velocity  ($v_i$) is treated as a nuisance parameter, and the true redshift is related to the true velocity from Eq.~\ref{eq:cz}. The regression problem therefore involves seven free parameters, namely $H_0$ and six true velocities, corresponding to each of the six megamaser  galaxy systems. P20 used  Bayesian inference for this analysis. They used uniform priors for all the free parameters and used {\tt dynesty}~\citep{dynesty} Nested sampler to sample the posterior constructed using the likelihood defined earlier. The marginalized value for $H_0$ obtained using this method is given by $H_0=73.9 \pm 3.0 $ km/sec/Mpc. Subsequently, different approaches were used in  correcting the peculiar velocities,   which do not modify $H_0$ by more than $1\sigma$. More details on this can be found in  P20.

\section{Bayesian estimate of $H_0$}
\label{sec:bayesian}
\begin{figure}[h]
    \centering
    \includegraphics[width=0.9\linewidth]{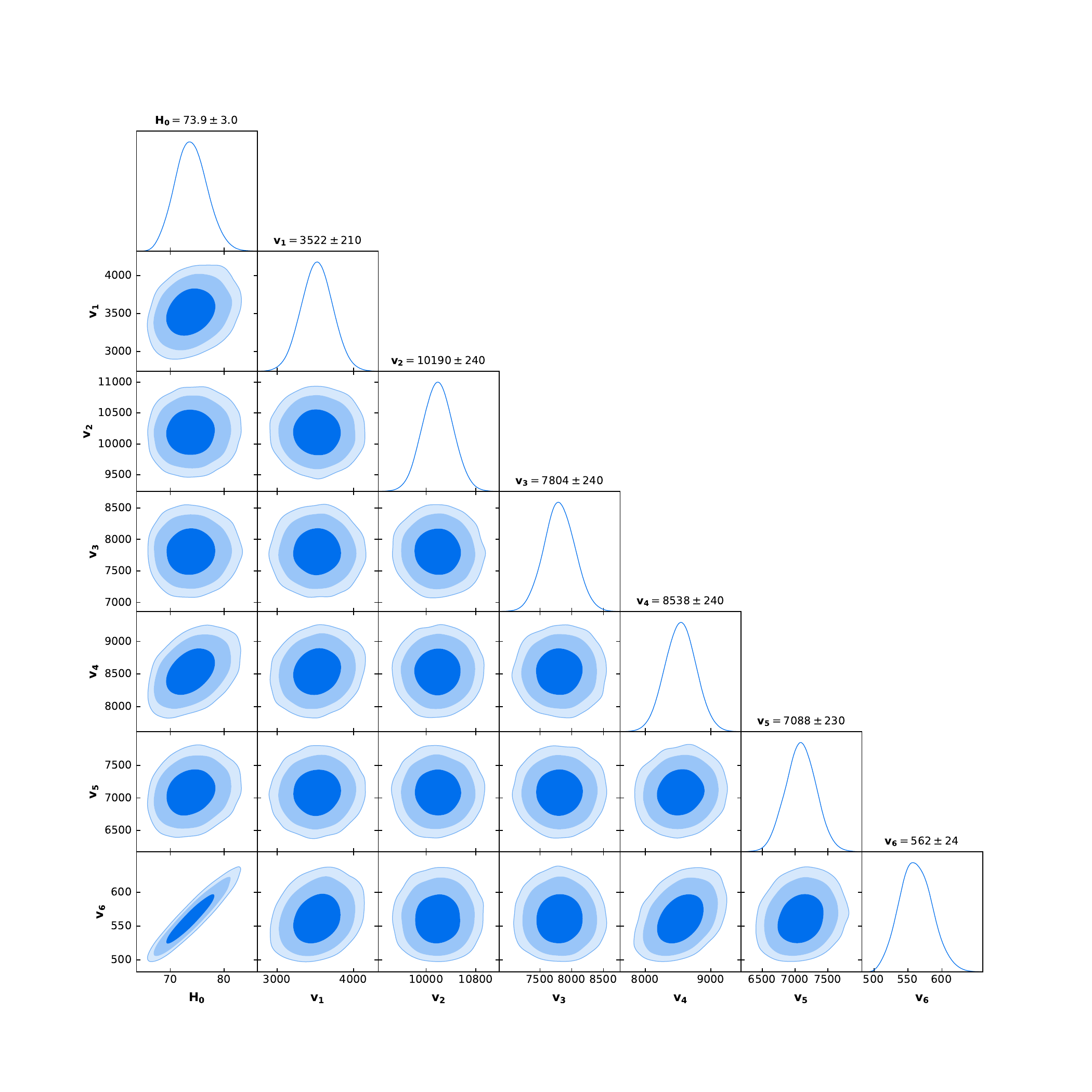}
    \caption{Marginalized 68\%, 95\%, and 99\% credible intervals for all the seven free parameters used in analysis of megamaser data. The marginalized $1\sigma$ central estimate of $H_0$ is given by $73.9\pm3.0$ km/sec/Mpc. \{$v1, v2, ... ,v6$\} correspond to the true velocities of UGC 3789, NGC 6264, NGC 6323, NGC 5765b, CGCG 074-064, and  NGC 4258, respectively.}
    \label{fig1}
\end{figure}

Before  proceeding to frequentist analysis, we independently reproduce the results in P20 using Bayesian regression, following the discussion in the previous section. For  our regression analysis, we define the unknown parameter vector by  \{$H_0$,$v_1$,$v_2$,$v_3$,$v_4$,$v_5$,$v_6$\}, where \{$v_1$, $v_2$,..,$v_6$\} denote the unknown true velocities of the six megamaser 
galaxy systems, which are treated as nuisance parameters. We use uniform priors for each of the true velocities given by $v_i \in \mathcal{U} (500,20000)$ in units of km/sec 
and $H_0 \in \mathcal{U} (50,200)$ in units of km/sec/Mpc. We sample the likelihood using {\tt emcee} sampler~\citep{emcee} and show the marginalized posteriors for each of the seven free parameters using the {\tt getdist} package~\citep{getdist}. The marginalized credible intervals for all the seven free parameters are shown in Fig.~\ref{fig1}. The marginalized $1\sigma$ estimate for $H_0$ is given by $H_0=73.9\pm3.0$ km/sec/Mpc, which is exactly in agreement with the value reported in P20. 
\rthis{Among all the corner plots, we only find a degeneracy between $H_0$ and $v_6$, corresponding to the galaxy NGC 4258,  because this galaxy has the smallest fractional error in distance of about 1.4\%. }
A comparison of this $H_0$ result with some other key measurements from CMB, Type Ia Supernova, as well as time-delay lensing can be found in Table~\ref{table4}. A more detailed compendium of all other $H_0$ measurements can be found in ~\citet{DiValentino22}.

\begin{table}[htbp!]
    \caption{\label{table4} Comparison of $H_0$ measurements from Megamasers with  a few other selected  probes along with the statistical significance of the discrepancy.}
    \centering
    \begin{tabular}{|c|c|c|}
        \hline
        \thead{Experiment} & \thead{$H_0$ (km/s/Mpc)} & \thead{Discrepancy} \\
        \hline
        Planck18 CMB (TT,TE,EE$+$low E$+$lensing)\tablefootnote{\cite{planck_2020}} & $67.36\pm0.54$ & $2.13 \sigma$ \\
        Pantheon$+$ Type Ia SNe\tablefootnote{\cite{riess_2022}} & $73.04\pm1.04$ & $0.2 \sigma$ \\
        TDCOSMO (Seven lens system)\tablefootnote{\cite{wong_2020}} & $74.2\pm1.6$ & $0.2 \sigma$ \\
        TDCOSMO$+$SLACS\tablefootnote{\cite{birrer_2020}} & $67.4^{+4.1}_{-3.2}$ & $1.4\sigma$ \\
        \hline
    \end{tabular}
\end{table}

\section{Frequentist estimate of $H_0$}
\label{sec:frequentist}
We now carry out a frequentist analysis of the megamaser data in order to estimate $H_0$, where we  use profile likelihood to deal with the six nuisance parameters. 
%More details on the basic principles behind profile likelihood can be found in ~\cite{Herold24} or our recent works~\citep{DesaiGanguly,Vyaas}, and references therein.

\begin{figure}
    \centering
    \includegraphics[width=0.75\linewidth]{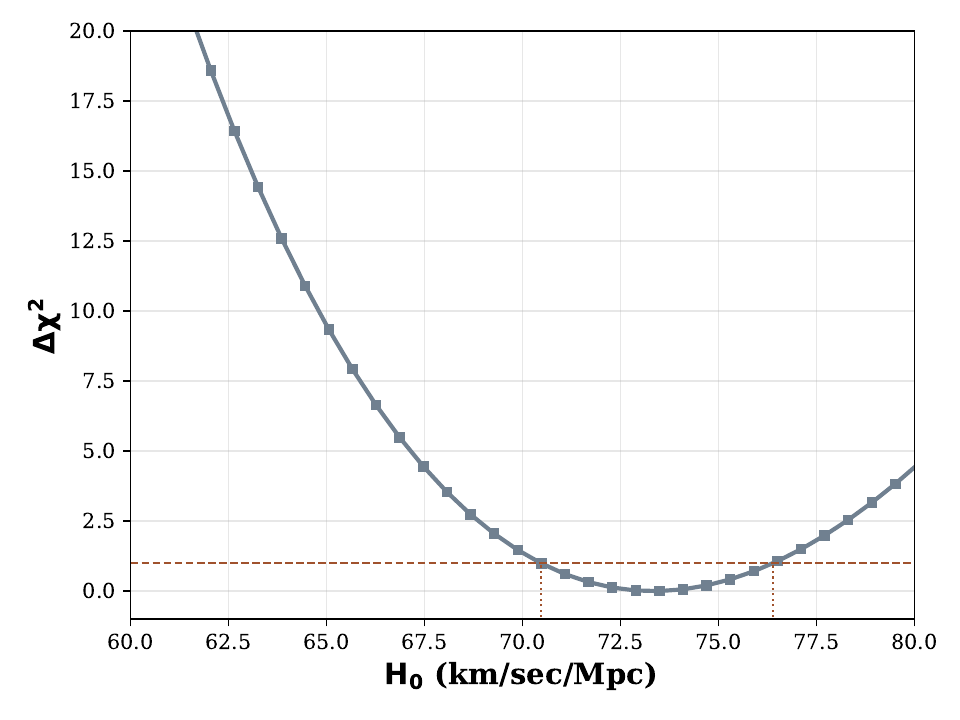}
    \caption{$\Delta \chi^2$ (cf. Eq.~\ref{eq:delchisq}) as a function of $H_0$ after using profile likelihood to deal with the six nuisance parameters. The dashed horizontal line corresponds to $\Delta \chi^2=1$ and the dotted line shows the corresponding $X$-intercept used to obtain our frequentist estimate of $H_0$. The $1\sigma$  estimate of $H_0$ is given by $73.5^{+3.0}_{-2.9}$ km/sec/Mpc.}
    \label{fig2}
\end{figure}

We first construct a uniformly spaced grid for $H_0$. For each fixed value of $H_0$ on this grid, we minimize $\chi^2$ in Eq.~\ref{eqn:ChiSquared} over the remaining six free parameters to obtain $\chi^2(H_0)$. This minimization for each fixed value of $H_0$ was done using the {\tt scipy.optimize.minimize} library in Python.  We tried six different minimization algorithms available in this function, namely  Nelder-Mead Simplex method, Powell, CG, BFGS, Newton-CG and L-BFGS-B  algorithm~\footnote{\url{https://docs.scipy.org/doc/scipy/reference/generated/scipy.optimize.minimize.html}}. We also tested minimization using the {\tt iminuit}\footnote{\url{https://scikit-hep.org/iminuit/}} library, which has previously been shown to be more robust than {\tt scipy}~\citep{Campeti}. However, for our minimization function, all the aforementioned algorithms return the same minimum value of $\chi^2$ at  each value of $H_0$. We then calculate the minimum $\chi^2$ among the different values of  $\chi^2 (H_0)$, and plot $\Delta \chi^2$ as a function of $H_0$ where 
\begin{equation}
\label{eq:delchisq}
\Delta \chi^2 = \chi^2 (H_0)- \chi^2_{min}
\end{equation}
This plot of $\Delta \chi^2$ as a function of $H_0$ can be found in Fig.~\ref{fig2}. According to Wilks' theorem, $\Delta\chi^2$ follows a $\chi^2$ distribution  for one degree of freedom ~\citep{Wilks1938}. Strictly speaking, Wilks' theorem only holds in the asymptotic limit of a large dataset. In case the best-fit value is close to the physical boundary, one uses the Feldman-Cousins prescription~\citep{FC}. However, since the minimum value obtained is far from the physical boundary, we use the Neyman prescription to get the central estimates for $H_0$~\citep{Neyman37}
To obtain the 68.3\% ($1\sigma$) confidence level estimates of $H_0$, we find  the $X$-intercept corresponding to $\Delta \chi^2=1.0$~\citep{NR}. The  $1\sigma$ central estimate of $H_0$ is then  given by  $H_0=73.5^{+3.0}_{-2.9}$ km/sec/Mpc. We can see that the best-fit values  agree with the Bayesian estimate    within $0.2\sigma$ and they also produce consistent confidence intervals.

Therefore, we have conducted a proof-of-principle application of profile likelihood by applying it to the problem of estimating  $H_0$ using data from the Megamaser Cosmology Project.
In this case, the results from both the frequentist and Bayesian estimates give statistically indistinguishable results. 
%whereas previously,   these methods  have led to  discordant results for some other regression problems in Astrophysics and Cosmology~\citep{Herold22,DesaiGanguly,Vyaas}.

\section{Comparison of results with $\Omega_m$ as a free parameter}
\label{sec:comparison}
\subsection{Bayesian Analysis}
\begin{figure}[H]
    \centering
    \includegraphics[width=0.9\linewidth]{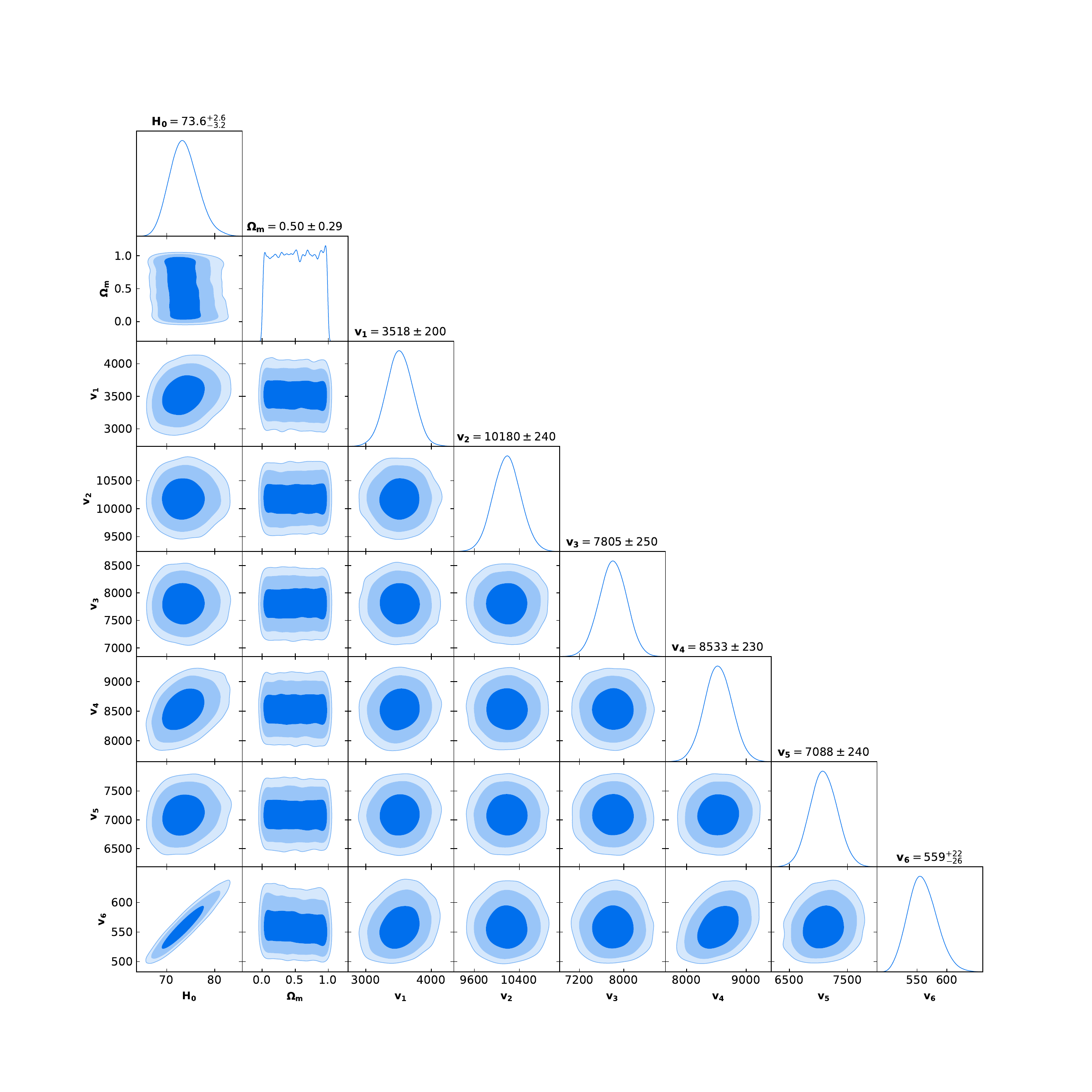}
    \caption{Marginalized 68\%, 95\%, and 99\% credible intervals for $H_0, \Omega_m$, and the true velocities $v_i$'s. The marginalized $1\sigma$ central estimate of $H_0$ is given by $73.6^{+2.6}_{-3.2}$ km/sec/Mpc. \{$v1, v2, ... ,v6$\} correspond to the true velocities of UGC 3789, NGC 6264, NGC 6323, NGC 5765b, CGCG 074-064, and  NGC 4258, respectively.}
    \label{fig3}
\end{figure}

Similar to Sect.~\ref{sec:bayesian}, we repeat the analysis, this time treating $\Omega_m$ as a free parameter with a prior range of $\mathcal{U}(0, 1)$.  We obtain  a value of $73.6^{+2.6}_{-3.2}$ km/sec/Mpc.  The corresponding corner plot can be found in Fig.~\ref{fig3}. Therefore,  
we do not observe any significant deviation in the value of $H_0$  (compared to when $\Omega_M$ was a fixed parameter). However, as shown in Fig.~\ref{fig3}, the posteriors on $\Omega_m$ remain broad, since the megamaser dataset does not constrain $\Omega_m$. This limited constraining power of the megamaser dataset prevents it from capturing the correlation between $H_0$ and $\Omega_m$. Consequently, from a Bayesian perspective, whether $\Omega_m$ is fixed or treated as a free parameter has little impact on the final $H_0$ results derived from the Megamaser data.

\subsection{Frequentist Analysis}
\begin{figure}[H]
    \centering
    \includegraphics[width=0.75\linewidth]{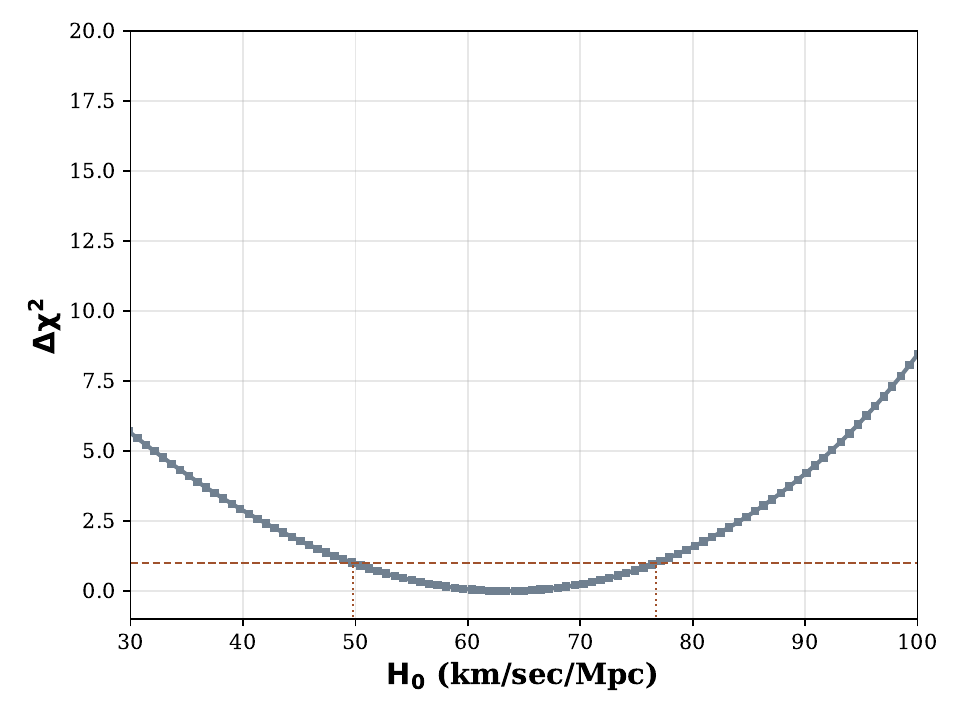}
    \caption{$\Delta \chi^2$ (cf. Eq.~\ref{eq:delchisq}) as a function of $H_0$ after using profile likelihood to deal with the seven nuisance parameters (including $\Omega_m$). The dashed horizontal line corresponds to $\Delta \chi^2=1$ and the dotted line shows the corresponding $X$-intercept used to obtain our frequentist estimate of $H_0$. In this case, $1\sigma$ estimate of $H_0$ is given by $63.4^{+13.3}_{-13.6}$ km/sec/Mpc.}
    \label{dchi2_variable_om}
\end{figure}

Similar to Sect.~\ref{sec:frequentist}, we repeat the analysis, this time treating $\Omega_m$ as a nuisance parameter. The corresponding $\Delta \chi^2$ curve as a function of $H_0$ can be found in Fig.~\ref{dchi2_variable_om}. The best-fit value of $H_0$ is given by $63.4^{+13.3}_{-13.6}$ km/sec/Mpc. Therefore, we find that although  $H_0$ is consistent with the Bayesian estimate to within $1\sigma$, the precision in the measurement gets degraded, with the $1\sigma$ uncertainty becoming five times larger compared to when $\Omega_m$ was fixed. However, we should point out  that the number of free parameters for a fixed $H_0$ is more than the number of data points. Therefore, this could be the main reason affecting the precision in the frequentist measurement of $H_0$ when $\Omega_m$ is also  kept free.

\section{Conclusions}
\label{sec:conc}
In this manuscript, we  have obtained an independent estimate of $H_0$ with data from the Megamaser Cosmology Project using frequentist inference. The main difference between the two methods is the treatment of nuisance parameters. In Bayesian inference, the nuisance parameters are dispensed with using marginalization, whereas in frequentist inference, profile likelihood is used for the same. 
This estimate of $H_0$ was first done using Bayesian inference in P20, which obtained a value of  $73.9 \pm 3.0$ km/sec/Mpc. This estimate of $H_0$ involved marginalization over six nuisance parameters, corresponding to the velocities of the six megamaser galaxy systems used for the analysis.
We first reproduced the result using Bayesian inference and independently obtained the value of $H_0=73.9\pm3.0$ km/sec/Mpc (cf. Fig.~\ref{fig1}), which agrees with the P20 estimate. We then redid the analysis using frequentist inference and obtained the value of $H_0=73.5^{+3.0}_{-2.9}$ km/sec/Mpc (cf. Fig.~\ref{fig2}). Therefore, the frequentist estimate of $H_0$ agrees with the result from Bayesian analysis within $0.2\sigma$. Furthermore,  both approaches yield  consistent confidence/credible intervals for fixed $\Omega_m$. thus implying that the results  are statistically indistinguishable.  
However, when we kept $\Omega_m$ as a free parameter, although  both the frequentist and Bayesian estimate of $H_0$  agreed with previous results, the precision in the profile likelihood based measurement of $H_0$ gets degraded by a factor of five compared to when $\Omega_m$ is fixed.

Therefore, this estimate of $H_0$ provides another proof of principle application of profile likelihood in dealing with nuisance parameters.

\section*{Acknowledgments}
SB would like to extend his gratitude to the University Grants Commission (UGC), Govt. of India for their continuous support through the Junior Research Fellowship. We are grateful to Dom Pesce for patiently explaining to us the  analysis method used in P20 and for very useful feedback on our manuscript. We also thank the anonymous referee for very constructive and useful comments on our manuscript.

\bibliographystyle{sn-basic}
\bibliography{main}

\begin{thebibliography}{46}
\providecommand{\natexlab}[1]{#1}
\providecommand{\url}[1]{{#1}}
\providecommand{\urlprefix}{URL }
\providecommand{\doi}[1]{\url{https://doi.org/#1}}
\providecommand{\eprint}[2][]{\url{#2}}
 \bibcommenthead

\bibitem[{Abdalla et~al.(2022)}]{tensionreview}
Abdalla E, et~al (2022) {Cosmology intertwined: A review of the particle physics, astrophysics, and cosmology associated with the cosmological tensions and anomalies}. JHEAp 34:49--211. \doi{10.1016/j.jheap.2022.04.002}, {\href{https://arxiv.org/abs/2203.06142}{{arXiv:2203.06142}}} {[astro-ph.CO]}

\bibitem[{Ade et~al.(2014)}]{Planck13}
Ade PAR, et~al (2014) {Planck intermediate results. XVI. Profile likelihoods for cosmological parameters}. Astron Astrophys 566:A54. \doi{10.1051/0004-6361/201323003}, {\href{https://arxiv.org/abs/1311.1657}{{arXiv:1311.1657}}} {[astro-ph.CO]}

\bibitem[{{Bethapudi} and {Desai}(2017)}]{Bethapudi}
{Bethapudi} S, {Desai} S (2017) {Median statistics estimates of Hubble and Newton's constants}. European Physical Journal Plus 132(2):78. \doi{10.1140/epjp/i2017-11390-3}, {\href{https://arxiv.org/abs/1701.01789}{{arXiv:1701.01789}}} {[astro-ph.CO]}

\bibitem[{{Birrer} et~al.(2020){Birrer}, {Shajib}, {Galan}, {Millon}, {Treu}, {Agnello}, {Auger}, {Chen}, {Christensen}, {Collett}, {Courbin}, {Fassnacht}, {Koopmans}, {Marshall}, {Park}, {Rusu}, {Sluse}, {Spiniello}, {Suyu}, {Wagner-Carena}, {Wong}, {Barnab{\`e}}, {Bolton}, {Czoske}, {Ding}, {Frieman}, and {Van de Vyvere}}]{birrer_2020}
{Birrer} S, {Shajib} AJ, {Galan} A, et~al (2020) {TDCOSMO. IV. Hierarchical time-delay cosmography - joint inference of the Hubble constant and galaxy density profiles}. \aap 643:A165. \doi{10.1051/0004-6361/202038861}, {\href{https://arxiv.org/abs/2007.02941}{{arXiv:2007.02941}}} {[astro-ph.CO]}

\bibitem[{{Braatz} et~al.(2008){Braatz}, {Reid}, {Greenhill}, {Condon}, {Lo}, {Henkel}, {Gugliucci}, and {Hao}}]{Braatz08}
{Braatz} JA, {Reid} MJ, {Greenhill} LJ, et~al (2008) {Investigating Dark Energy with Observations of H$_{2}$O Megamasers}. In: {Bridle} AH, {Condon} JJ, {Hunt} GC (eds) Frontiers of Astrophysics: A Celebration of NRAO's 50th Anniversary, p 103

\bibitem[{{Campeti} and {Komatsu}(2022)}]{Campeti}
{Campeti} P, {Komatsu} E (2022) {New Constraint on the Tensor-to-scalar Ratio from the Planck and BICEP/Keck Array Data Using the Profile Likelihood}. \apj 941(2):110. \doi{10.3847/1538-4357/ac9ea3}, {\href{https://arxiv.org/abs/2205.05617}{{arXiv:2205.05617}}} {[astro-ph.CO]}

\bibitem[{{Carrilho} et~al.(2023){Carrilho}, {Moretti}, and {Pourtsidou}}]{Pedro}
{Carrilho} P, {Moretti} C, {Pourtsidou} A (2023) {Cosmology with the EFTofLSS and BOSS: dark energy constraints and a note on priors}. \jcap 2023(1):028. \doi{10.1088/1475-7516/2023/01/028}, {\href{https://arxiv.org/abs/2207.14784}{{arXiv:2207.14784}}} {[astro-ph.CO]}

\bibitem[{{Colg{\'a}in} et~al.(2024){Colg{\'a}in}, {Pourojaghi}, and {Sheikh-Jabbari}}]{Colgain24}
{Colg{\'a}in} E{\'O}, {Pourojaghi} S, {Sheikh-Jabbari} MM (2024) {Implications of DES 5YR SNe Dataset for $\Lambda$CDM}. arXiv e-prints arXiv:2406.06389. \doi{10.48550/arXiv.2406.06389}, {\href{https://arxiv.org/abs/2406.06389}{{arXiv:2406.06389}}} {[astro-ph.CO]}

\bibitem[{{Cowan}(2013)}]{Cowan13}
{Cowan} G (2013) {Statistics for Searches at the LHC}. arXiv e-prints arXiv:1307.2487. \doi{10.48550/arXiv.1307.2487}, {\href{https://arxiv.org/abs/1307.2487}{{arXiv:1307.2487}}} {[hep-ex]}

\bibitem[{{Desai} and {Ganguly}(2025)}]{DesaiGanguly}
{Desai} S, {Ganguly} S (2025) {Constraint on Lorentz invariance violation for spectral lag transition in GRB 160625B using profile likelihood}. European Physical Journal C 85(3):290. \doi{10.1140/epjc/s10052-025-14016-0}, {\href{https://arxiv.org/abs/2411.09248}{{arXiv:2411.09248}}} {[astro-ph.HE]}

\bibitem[{{Di Valentino} et~al.(2021){Di Valentino}, {Mena}, {Pan}, {Visinelli}, {Yang}, {Melchiorri}, {Mota}, {Riess}, and {Silk}}]{DiValentino22}
{Di Valentino} E, {Mena} O, {Pan} S, et~al (2021) {In the realm of the Hubble tension-a review of solutions}. Classical and Quantum Gravity 38(15):153001. \doi{10.1088/1361-6382/ac086d}, {\href{https://arxiv.org/abs/2103.01183}{{arXiv:2103.01183}}} {[astro-ph.CO]}

\bibitem[{{Feldman} and {Cousins}(1998)}]{FC}
{Feldman} GJ, {Cousins} RD (1998) {Unified approach to the classical statistical analysis of small signals}. \prd 57(7):3873--3889. \doi{10.1103/PhysRevD.57.3873}, {\href{https://arxiv.org/abs/physics/9711021}{{arXiv:physics/9711021}}} {[physics.data-an]}

\bibitem[{{Foreman-Mackey} et~al.(2013){Foreman-Mackey}, {Hogg}, {Lang}, and {Goodman}}]{emcee}
{Foreman-Mackey} D, {Hogg} DW, {Lang} D, et~al (2013) {emcee: The MCMC Hammer}. \pasp 125(925):306. \doi{10.1086/670067}, {\href{https://arxiv.org/abs/1202.3665}{{arXiv:1202.3665}}} {[astro-ph.IM]}

\bibitem[{{Freedman}(2021)}]{Freedman21}
{Freedman} WL (2021) {Measurements of the Hubble Constant: Tensions in Perspective}. \apj 919(1):16. \doi{10.3847/1538-4357/ac0e95}, {\href{https://arxiv.org/abs/2106.15656}{{arXiv:2106.15656}}} {[astro-ph.CO]}

\bibitem[{G\'omez-Valent(2022)}]{Adria}
G\'omez-Valent A (2022) {Fast test to assess the impact of marginalization in Monte~Carlo analyses and its application to cosmology}. Phys Rev D 106(6):063506. \doi{10.1103/PhysRevD.106.063506}, {\href{https://arxiv.org/abs/2203.16285}{{arXiv:2203.16285}}} {[astro-ph.CO]}

\bibitem[{{Gsponer} et~al.(2024){Gsponer}, {Zhao}, {Donald-McCann}, {Bacon}, {Koyama}, {Crittenden}, {Simon}, and {Mueller}}]{Bacon23}
{Gsponer} R, {Zhao} R, {Donald-McCann} J, et~al (2024) {Cosmological constraints on early dark energy from the full shape analysis of eBOSS DR16}. \mnras 530(3):3075--3099. \doi{10.1093/mnras/stae992}, {\href{https://arxiv.org/abs/2312.01977}{{arXiv:2312.01977}}} {[astro-ph.CO]}

\bibitem[{{Hadzhiyska} et~al.(2023){Hadzhiyska}, {Wolz}, {Azzoni}, {Alonso}, {Garc{\'\i}a-Garc{\'\i}a}, {Ruiz-Zapatero}, and {Slosar}}]{Slosar23}
{Hadzhiyska} B, {Wolz} K, {Azzoni} S, et~al (2023) {Cosmology with 6 parameters in the Stage-IV era: efficient marginalisation over nuisance parameters}. The Open Journal of Astrophysics 6:23. \doi{10.21105/astro.2301.11895}, {\href{https://arxiv.org/abs/2301.11895}{{arXiv:2301.11895}}} {[astro-ph.CO]}

\bibitem[{{Hamann}(2012)}]{Hamann12}
{Hamann} J (2012) {Evidence for extra radiation? Profile likelihood versus Bayesian posterior}. \jcap 2012(3):021. \doi{10.1088/1475-7516/2012/03/021}, {\href{https://arxiv.org/abs/1110.4271}{{arXiv:1110.4271}}} {[astro-ph.CO]}

\bibitem[{{Herold} and {Ferreira}(2023)}]{HeroldFerreira}
{Herold} L, {Ferreira} EGM (2023) {Resolving the Hubble tension with early dark energy}. \prd 108(4):043513. \doi{10.1103/PhysRevD.108.043513}, {\href{https://arxiv.org/abs/2210.16296}{{arXiv:2210.16296}}} {[astro-ph.CO]}

\bibitem[{{Herold} et~al.(2022){Herold}, {Ferreira}, and {Komatsu}}]{Herold22}
{Herold} L, {Ferreira} EGM, {Komatsu} E (2022) {New Constraint on Early Dark Energy from Planck and BOSS Data Using the Profile Likelihood}. \apjl 929(1):L16. \doi{10.3847/2041-8213/ac63a3}, {\href{https://arxiv.org/abs/2112.12140}{{arXiv:2112.12140}}} {[astro-ph.CO]}

\bibitem[{{Herold} et~al.(2025){Herold}, {Ferreira}, and {Heinrich}}]{Herold24}
{Herold} L, {Ferreira} EGM, {Heinrich} L (2025) {Profile likelihoods in cosmology: When, why, and how illustrated with \ensuremath{\Lambda}, massive neutrinos, and dark energy}. \prd 111(8):083504. \doi{10.1103/PhysRevD.111.083504}, {\href{https://arxiv.org/abs/2408.07700}{{arXiv:2408.07700}}} {[astro-ph.CO]}

\bibitem[{{Herrnstein} et~al.(1999){Herrnstein}, {Moran}, {Greenhill}, {Diamond}, {Inoue}, {Nakai}, {Miyoshi}, {Henkel}, and {Riess}}]{Hern}
{Herrnstein} JR, {Moran} JM, {Greenhill} LJ, et~al (1999) {A geometric distance to the galaxy NGC4258 from orbital motions in a nuclear gas disk}. \nat 400(6744):539--541. \doi{10.1038/22972}, {\href{https://arxiv.org/abs/astro-ph/9907013}{{arXiv:astro-ph/9907013}}} {[astro-ph]}

\bibitem[{{Kamionkowski} and {Riess}(2023)}]{KamionkowskiRiess}
{Kamionkowski} M, {Riess} AG (2023) {The Hubble Tension and Early Dark Energy}. Annual Review of Nuclear and Particle Science 73:153--180. \doi{10.1146/annurev-nucl-111422-024107}, {\href{https://arxiv.org/abs/2211.04492}{{arXiv:2211.04492}}} {[astro-ph.CO]}

\bibitem[{{Karwal} et~al.(2024){Karwal}, {Patel}, {Bartlett}, {Poulin}, {Smith}, and {Pfeffer}}]{Karwal24}
{Karwal} T, {Patel} Y, {Bartlett} A, et~al (2024) {Procoli: Profiles of cosmological likelihoods}. arXiv e-prints arXiv:2401.14225. \doi{10.48550/arXiv.2401.14225}, {\href{https://arxiv.org/abs/2401.14225}{{arXiv:2401.14225}}} {[astro-ph.CO]}

\bibitem[{{Lewis}(2019)}]{getdist}
{Lewis} A (2019) {GetDist: a Python package for analysing Monte Carlo samples}. arXiv e-prints arXiv:1910.13970. \doi{10.48550/arXiv.1910.13970}, {\href{https://arxiv.org/abs/1910.13970}{{arXiv:1910.13970}}} {[astro-ph.IM]}

\bibitem[{Neyman(1937)}]{Neyman37}
Neyman J (1937) Outline of a theory of statistical estimation based on the classical theory of probability. Philosophical Transactions of the Royal Society of London Series A, Mathematical and Physical Sciences 236(767):333--380

\bibitem[{{Nielsen} et~al.(2016){Nielsen}, {Guffanti}, and {Sarkar}}]{Sarkar16}
{Nielsen} JT, {Guffanti} A, {Sarkar} S (2016) {Marginal evidence for cosmic acceleration from Type Ia supernovae}. Scientific Reports 6:35596. \doi{10.1038/srep35596}, {\href{https://arxiv.org/abs/1506.01354}{{arXiv:1506.01354}}} {[astro-ph.CO]}

\bibitem[{{Pesce} et~al.(2020{\natexlab{a}}){Pesce}, {Braatz}, {Reid}, {Condon}, {Gao}, {Henkel}, {Kuo}, {Lo}, and {Zhao}}]{Pesce20}
{Pesce} DW, {Braatz} JA, {Reid} MJ, et~al (2020{\natexlab{a}}) {The Megamaser Cosmology Project. XI. A Geometric Distance to CGCG 074-064}. \apj 890(2):118. \doi{10.3847/1538-4357/ab6bcd}, {\href{https://arxiv.org/abs/2001.04581}{{arXiv:2001.04581}}} {[astro-ph.GA]}

\bibitem[{{Pesce} et~al.(2020{\natexlab{b}}){Pesce}, {Braatz}, {Reid}, {Riess}, {Scolnic}, {Condon}, {Gao}, {Henkel}, {Impellizzeri}, {Kuo}, and {Lo}}]{Pesce}
{Pesce} DW, {Braatz} JA, {Reid} MJ, et~al (2020{\natexlab{b}}) {The Megamaser Cosmology Project. XIII. Combined Hubble Constant Constraints}. \apjl 891(1):L1. \doi{10.3847/2041-8213/ab75f0}, {\href{https://arxiv.org/abs/2001.09213}{{arXiv:2001.09213}}} {[astro-ph.CO]}

\bibitem[{{Planck Collaboration} et~al.(2020){Planck Collaboration}, {Aghanim}, {Akrami}, {Ashdown}, {Aumont}, {Baccigalupi}, {Ballardini}, {Banday}, {Barreiro}, {Bartolo}, {Basak}, {Battye}, {Benabed}, {Bernard}, {Bersanelli}, {Bielewicz}, {Bock}, {Bond}, {Borrill}, {Bouchet}, {Boulanger}, {Bucher}, {Burigana}, {Butler}, {Calabrese}, {Cardoso}, {Carron}, {Challinor}, {Chiang}, {Chluba}, {Colombo}, {Combet}, {Contreras}, {Crill}, {Cuttaia}, {de Bernardis}, {de Zotti}, {Delabrouille}, {Delouis}, {Di Valentino}, {Diego}, {Dor{\'e}}, {Douspis}, {Ducout}, {Dupac}, {Dusini}, {Efstathiou}, {Elsner}, {En{\ss}lin}, {Eriksen}, {Fantaye}, {Farhang}, {Fergusson}, {Fernandez-Cobos}, {Finelli}, {Forastieri}, {Frailis}, {Fraisse}, {Franceschi}, {Frolov}, {Galeotta}, {Galli}, {Ganga}, {G{\'e}nova-Santos}, {Gerbino}, {Ghosh}, {Gonz{\'a}lez-Nuevo}, {G{\'o}rski}, {Gratton}, {Gruppuso}, {Gudmundsson}, {Hamann}, {Handley}, {Hansen}, {Herranz}, {Hildebrandt}, {Hivon}, {Huang}, {Jaffe}, {Jones}, {Karakci}, {Keih{\"a}nen},
  {Keskitalo}, {Kiiveri}, {Kim}, {Kisner}, {Knox}, {Krachmalnicoff}, {Kunz}, {Kurki-Suonio}, {Lagache}, {Lamarre}, {Lasenby}, {Lattanzi}, {Lawrence}, {Le Jeune}, {Lemos}, {Lesgourgues}, {Levrier}, {Lewis}, {Liguori}, {Lilje}, {Lilley}, {Lindholm}, {L{\'o}pez-Caniego}, {Lubin}, {Ma}, {Mac{\'\i}as-P{\'e}rez}, {Maggio}, {Maino}, {Mandolesi}, {Mangilli}, {Marcos-Caballero}, {Maris}, {Martin}, {Martinelli}, {Mart{\'\i}nez-Gonz{\'a}lez}, {Matarrese}, {Mauri}, {McEwen}, {Meinhold}, {Melchiorri}, {Mennella}, {Migliaccio}, {Millea}, {Mitra}, {Miville-Desch{\^e}nes}, {Molinari}, {Montier}, {Morgante}, {Moss}, {Natoli}, {N{\o}rgaard-Nielsen}, {Pagano}, {Paoletti}, {Partridge}, {Patanchon}, {Peiris}, {Perrotta}, {Pettorino}, {Piacentini}, {Polastri}, {Polenta}, {Puget}, {Rachen}, {Reinecke}, {Remazeilles}, {Renzi}, {Rocha}, {Rosset}, {Roudier}, {Rubi{\~n}o-Mart{\'\i}n}, {Ruiz-Granados}, {Salvati}, {Sandri}, {Savelainen}, {Scott}, {Shellard}, {Sirignano}, {Sirri}, {Spencer}, {Sunyaev}, {Suur-Uski}, {Tauber}, {Tavagnacco},
  {Tenti}, {Toffolatti}, {Tomasi}, {Trombetti}, {Valenziano}, {Valiviita}, {Van Tent}, {Vibert}, {Vielva}, {Villa}, {Vittorio}, {Wandelt}, {Wehus}, {White}, {White}, {Zacchei}, and {Zonca}}]{planck_2020}
{Planck Collaboration}, {Aghanim} N, {Akrami} Y, et~al (2020) {Planck 2018 results. VI. Cosmological parameters}. \aap 641:A6. \doi{10.1051/0004-6361/201833910}, {\href{https://arxiv.org/abs/1807.06209}{{arXiv:1807.06209}}} {[astro-ph.CO]}

\bibitem[{{Press} et~al.(1992){Press}, {Teukolsky}, {Vetterling}, and {Flannery}}]{NR}
{Press} WH, {Teukolsky} SA, {Vetterling} WT, et~al (1992) {Numerical recipes in FORTRAN. The art of scientific computing}

\bibitem[{{Ramakrishnan} and {Desai}(2025)}]{Vyaas}
{Ramakrishnan} V, {Desai} S (2025) {Constraints on Lorentz Invariance Violation from Gamma-ray Burst rest-frame spectral lags using Profile Likelihood}. arXiv e-prints arXiv:2502.00805. \doi{10.48550/arXiv.2502.00805}, {\href{https://arxiv.org/abs/2502.00805}{{arXiv:2502.00805}}} {[astro-ph.HE]}

\bibitem[{{Reid} et~al.(2009){Reid}, {Braatz}, {Condon}, {Greenhill}, {Henkel}, and {Lo}}]{Reid09}
{Reid} MJ, {Braatz} JA, {Condon} JJ, et~al (2009) {The Megamaser Cosmology Project. I. Very Long Baseline Interferometric Observations of UGC 3789}. \apj 695(1):287--291. \doi{10.1088/0004-637X/695/1/287}, {\href{https://arxiv.org/abs/0811.4345}{{arXiv:0811.4345}}} {[astro-ph]}

\bibitem[{{Reid} et~al.(2019){Reid}, {Pesce}, and {Riess}}]{Reid19}
{Reid} MJ, {Pesce} DW, {Riess} AG (2019) {An Improved Distance to NGC 4258 and Its Implications for the Hubble Constant}. \apjl 886(2):L27. \doi{10.3847/2041-8213/ab552d}, {\href{https://arxiv.org/abs/1908.05625}{{arXiv:1908.05625}}} {[astro-ph.GA]}

\bibitem[{{Riess} et~al.(2022){Riess}, {Yuan}, {Macri}, {Scolnic}, {Brout}, {Casertano}, {Jones}, {Murakami}, {Anand}, {Breuval}, {Brink}, {Filippenko}, {Hoffmann}, {Jha}, {D'arcy Kenworthy}, {Mackenty}, {Stahl}, and {Zheng}}]{riess_2022}
{Riess} AG, {Yuan} W, {Macri} LM, et~al (2022) {A Comprehensive Measurement of the Local Value of the Hubble Constant with 1 km s$^{-1}$ Mpc$^{-1}$ Uncertainty from the Hubble Space Telescope and the SH0ES Team}. \apjl 934(1):L7. \doi{10.3847/2041-8213/ac5c5b}, {\href{https://arxiv.org/abs/2112.04510}{{arXiv:2112.04510}}} {[astro-ph.CO]}

\bibitem[{{Sah} et~al.(2024){Sah}, {Rameez}, {Sarkar}, and {Tsagas}}]{Sah24}
{Sah} A, {Rameez} M, {Sarkar} S, et~al (2024) {Anisotropy in Pantheon+ supernovae}. arXiv e-prints arXiv:2411.10838. \doi{10.48550/arXiv.2411.10838}, {\href{https://arxiv.org/abs/2411.10838}{{arXiv:2411.10838}}} {[astro-ph.CO]}

\bibitem[{{Shah} et~al.(2021){Shah}, {Lemos}, and {Lahav}}]{Shah21}
{Shah} P, {Lemos} P, {Lahav} O (2021) {A buyer's guide to the Hubble constant}. \aapr 29(1):9. \doi{10.1007/s00159-021-00137-4}, {\href{https://arxiv.org/abs/2109.01161}{{arXiv:2109.01161}}} {[astro-ph.CO]}

\bibitem[{{Sharma}(2017)}]{Sanjib}
{Sharma} S (2017) {Markov Chain Monte Carlo Methods for Bayesian Data Analysis in Astronomy}. \araa 55(1):213--259. \doi{10.1146/annurev-astro-082214-122339}, {\href{https://arxiv.org/abs/1706.01629}{{arXiv:1706.01629}}} {[astro-ph.IM]}

\bibitem[{{Smith} et~al.(2021){Smith}, {Poulin}, {Bernal}, {Boddy}, {Kamionkowski}, and {Murgia}}]{Smith21}
{Smith} TL, {Poulin} V, {Bernal} JL, et~al (2021) {Early dark energy is not excluded by current large-scale structure data}. \prd 103(12):123542. \doi{10.1103/PhysRevD.103.123542}, {\href{https://arxiv.org/abs/2009.10740}{{arXiv:2009.10740}}} {[astro-ph.CO]}

\bibitem[{{Speagle}(2020)}]{dynesty}
{Speagle} JS (2020) {DYNESTY: a dynamic nested sampling package for estimating Bayesian posteriors and evidences}. \mnras 493(3):3132--3158. \doi{10.1093/mnras/staa278}, {\href{https://arxiv.org/abs/1904.02180}{{arXiv:1904.02180}}} {[astro-ph.IM]}

\bibitem[{{Trotta}(2008)}]{Trotta08}
{Trotta} R (2008) {Bayes in the sky: Bayesian inference and model selection in cosmology}. Contemporary Physics 49(2):71--104. \doi{10.1080/00107510802066753}, {\href{https://arxiv.org/abs/0803.4089}{{arXiv:0803.4089}}} {[astro-ph]}

\bibitem[{{Vagnozzi}(2023)}]{Vagnozzi}
{Vagnozzi} S (2023) {Seven Hints That Early-Time New Physics Alone Is Not Sufficient to Solve the Hubble Tension}. Universe 9(9):393. \doi{10.3390/universe9090393}, {\href{https://arxiv.org/abs/2308.16628}{{arXiv:2308.16628}}} {[astro-ph.CO]}

\bibitem[{{Verde} et~al.(2019){Verde}, {Treu}, and {Riess}}]{Verde}
{Verde} L, {Treu} T, {Riess} AG (2019) {Tensions between the early and late Universe}. Nature Astronomy 3:891--895. \doi{10.1038/s41550-019-0902-0}, {\href{https://arxiv.org/abs/1907.10625}{{arXiv:1907.10625}}} {[astro-ph.CO]}

\bibitem[{{Verde} et~al.(2024){Verde}, {Sch{\"o}neberg}, and {Gil-Mar{\'\i}n}}]{Verde24}
{Verde} L, {Sch{\"o}neberg} N, {Gil-Mar{\'\i}n} H (2024) {A Tale of Many H $_{0}$}. \araa 62(1):287--331. \doi{10.1146/annurev-astro-052622-033813}, {\href{https://arxiv.org/abs/2311.13305}{{arXiv:2311.13305}}} {[astro-ph.CO]}

\bibitem[{Wilks(1938)}]{Wilks1938}
Wilks SS (1938) The large-sample distribution of the likelihood ratio for testing composite hypotheses. The annals of mathematical statistics 9(1):60--62

\bibitem[{{Wong} et~al.(2020){Wong}, {Suyu}, {Chen}, {Rusu}, {Millon}, {Sluse}, {Bonvin}, {Fassnacht}, {Taubenberger}, {Auger}, {Birrer}, {Chan}, {Courbin}, {Hilbert}, {Tihhonova}, {Treu}, {Agnello}, {Ding}, {Jee}, {Komatsu}, {Shajib}, {Sonnenfeld}, {Blandford}, {Koopmans}, {Marshall}, and {Meylan}}]{wong_2020}
{Wong} KC, {Suyu} SH, {Chen} GCF, et~al (2020) {H0LiCOW - XIII. A 2.4 per cent measurement of H$_{0}$ from lensed quasars: 5.3{\ensuremath{\sigma}} tension between early- and late-Universe probes}. \mnras 498(1):1420--1439. \doi{10.1093/mnras/stz3094}, {\href{https://arxiv.org/abs/1907.04869}{{arXiv:1907.04869}}} {[astro-ph.CO]}

\end{thebibliography}
\end{document}